\documentclass[aps,prb,twocolumn,amsmath,amssymb,superscriptaddress,floatfix]{revtex4}
\usepackage{graphicx}
\usepackage{bm}
\usepackage[usenames]{color}
\bibstyle{apsrev.bib}

\newcommand{\be}{\begin{equation}}
\newcommand{\ee}{\end{equation}}
\newcommand{\beqn}{\begin{eqnarray}}
\newcommand{\eeqn}{\end{eqnarray}}

\begin{document}

\title{Corner contribution to percolation cluster numbers in three dimensions}

\author{Istv\'an A. Kov\'acs}
\email{kovacs.istvan@wigner.mta.hu}
\affiliation{Wigner Research Centre, Institute for Solid State Physics and Optics, H-1525 Budapest, P.O.Box 49, Hungary}
\affiliation{Institute of Theoretical Physics, Szeged University, H-6720 Szeged, Hungary}
\author{Ferenc Igl\'oi}
\email{igloi.ferenc@wigner.mta.hu}
\affiliation{Wigner Research Centre, Institute for Solid State Physics and Optics, H-1525 Budapest, P.O.Box 49, Hungary}
\affiliation{Institute of Theoretical Physics, Szeged University, H-6720 Szeged, Hungary}
\date{\today}

\begin{abstract}
In three-dimensional critical percolation we study numerically the number of clusters, $N_{\Gamma}$, which intersect a
given subset of bonds, $\Gamma$. If $\Gamma$ represents the interface between a subsystem and the environment, then $N_{\Gamma}$ is
related to the entanglement entropy of the critical diluted quantum Ising model. Due to corners in $\Gamma$ there are
singular corrections to $N_{\Gamma}$, which scale as $b_{\Gamma} \ln L_{\Gamma}$, $L_{\Gamma}$ being the linear size
of $\Gamma$ and the prefactor, $b_{\Gamma}$, is found to be universal. This result indicates that logarithmic
finite-size corrections exist in the free-energy of three-dimensional critical systems.
\end{abstract}

\maketitle
\section{Introduction}
\label{sec:intr}
Percolation is a basic problem of geometrical critical phenomena, when 
sites or bonds of a regular lattice are independently open
with a probability $p$ and one is interested in the statistics of clusters of connected
sites\cite{stauffer}. Percolation is most studied in two dimensions ($2d$), in which case at the critical point
in the scaling limit the system is conformally invariant\cite{smirnov}. At the critical point many exact results are available
about the order-parameter profiles, correlation functions\cite{dotsenko_fateev}, crossing probabilities\cite{crossing}, etc.
through conformal invariance\cite{conf_inv} and some of those have been subsequently derived by rigorous
mathematical methods, such as by Schramm-Loewner evolution\cite{sle}.

Conformal invariance has also predictions about the finite-size corrections to the total number
of clusters in critical $2d$ percolation: the corner contribution is logarithmically divergent
and its prefactor is universal\cite{kovacs}. This prefactor has been calculated analytically through the application of the
Cardy-Peschel formula\cite{cardypeschel} and the results have been checked trough Monte-Carlo simulations. Very recently
these investigations have been extended to the Fortuin-Kasteleyn and spin clusters in the $Q \le 4$
state Potts model\cite{kovacs14} (percolation being recovered in the limit $Q \to 1$).

Finite-size corrections to the total number of clusters are of importance in three dimensions ($3d$), too, due
to relations to another physical problems. First, we mention the question about finite-size
scaling of the free-energy in $3d$ critical percolation, or more generally in $3d$
critical systems\cite{privman_fisher}. In this case there are no predictions from (conformal) field theory and an
analogue of the Cardy-Peschel formula does not exist. However, according to scaling theory the logarithm in
two-dimensions can be interpreted in terms of singular and non-singular contributions in the free energy and
according to this argument similar terms can also arise from corners in three dimensions\cite{privman}. To our knowledge
this type of corner corrections have not been verified till now. Our second related problem is the entanglement
entropy of the diluted quantum Ising model, see in Refs.\cite{lin07,yu07} and in the short description in Sec.\ref{sec:entropy}.
If there are logarithmic corrections to the entanglement entropy in the $3d$ model due to corners is an interesting
open question.

In this paper we continue and extend our previous $2d$ investigations\cite{kovacs} and study the finite-size behavior
of the critical percolation cluster numbers in $3d$ and in particular we investigate the corrections
coming from corners.
To be specific we consider bond percolation in a $3d$ simple cubic lattice and denote by $\Gamma$ a
subset of bonds. Initially we take this to be the surface of a cubic subsystem of edge $L_{\Gamma}$.
We are interested in the number of clusters, $N_{\Gamma}$,
which intersect $\Gamma$, in the scaling limit when $\Gamma$ is large but still much smaller than the total size of the system.
$N_{\Gamma}$ is closely related to the
entanglement entropy of the bond-diluted quantum Ising model which is defined
in the same lattice and $\Gamma$ represents the interface, which separates a subsystem from the
rest of the system.

Our paper is organized as follows.
In Sec.\ref{sec:Potts} we recapitulate known results. $N_{\Gamma}$ is expressed in the Potts model representation and
its relation with the entanglement entropy of the bond-diluted quantum Ising model is explained.
In the main part of the paper in Sec.\ref{sec:numerical} we calculate $N_{\Gamma}$ by large scale Monte Carlo simulations for different
forms of $\Gamma$. We concentrate on cube subsystems, but consider another type of contours as well, such as squares,
line segments and crosses. These calculations are made both for bond
and site percolation and we demonstrate the universality of the corner contribution. In Sec.\ref{sec:disc} our paper is closed
by a discussion.

\section{Reminder: percolation cluster numbers}
\label{sec:Potts}

\subsection{Potts model representation}

Bond percolation is equivalent to the $Q \to 1$ limit of the $Q$-state Potts model\cite{wu}, 
defined on a lattice with sites $i=1,2,\dots,n$ and $m$ nearest neighbor bonds. The partition sum
of the model, $Z(Q)$, is conveniently written in terms of the so called Fortuin-Kasteleyn
clusters\cite{Fortuin-Kasteleyn}, denoted by $F$. For a given element of $F$ there are $N_{\mathrm{tot}}(F) \le n$
connected components and $M(F) \le m$ occupied bonds and the partition function reads as:
\beqn
Z(Q) \sim \sum_{F} Q^{N_{\mathrm{tot}}(F)} p^{M(F)} {(1-p)}^{m-M(F)}\;.
\label{Z}
\eeqn
In this random cluster representation $Q$ is a real parameter, the bond occupation probability is $p=1-e^{-K}$,
$K$ being the reduced coupling and percolation is recovered in the $Q \to 1$ limit.
The mean total number of clusters in percolation is:
\be
\left\langle N_{\mathrm{tot}}\right\rangle=\left. \dfrac{\partial \ln Z(Q)}{\partial Q}\right|_{Q=1}\;.
\label{N_tot}
\ee
If we fix all spins on $\Gamma$ (in state $1$,
say), but leave the couplings unchanged
%
%
this relation is modified as:
\be
\left\langle N_\mathrm{tot} - N_{\Gamma} \right\rangle=\left.
\dfrac{\partial \ln Z_{\Gamma}(Q)}{\partial Q} \right|_{Q=1}\;,
\label{N_tot-N_Gamma}
\ee
where $N_{\Gamma}$ is the number of clusters which intersect $\Gamma$\cite{touching}.
At the critical point, $p=p_c$, we can write:
\beqn
\ln Z(Q)-\ln Z_{\Gamma}(Q)&\sim& S_{\Gamma} f_{\mathrm s}(Q) + L_{\Gamma} f_{\mathrm e}(Q)+ f_{\mathrm c}(Q),
\label{log_Z}
\eeqn
where $S_{\Gamma}$ and $L_{\Gamma}$ is the total area and the linear
extension of $\Gamma$, and
$f_{\mathrm s}$ and $f_{\mathrm e}$ are the surface and edge free-energy densities, respectively,
which are non-universal quantities. The last term in Eq.~\eqref{log_Z}, $f_{\mathrm c}(Q)$, represents
the corner contribution, which is known to be singular in $2d$\cite{cardypeschel}: $f_{\mathrm c}(Q)= C_{\Gamma}(Q) \ln L_{\Gamma}$.
Together with Eqs.~\eqref{N_tot} and \eqref{N_tot-N_Gamma}
we hence obtain:
\be
\left\langle N_{\Gamma} \right\rangle=-f'_{\mathrm s}(1)S_{\Gamma}-f'_{\mathrm e}(1)L_{\Gamma}-f'_{\mathrm c}(1) \;,
\label{N_Gamma}
\ee
Here we study this problem in $3d$ and we are interested in the behavior of the corner contribution.
We shall show, that like in $2d$ at the critical point $f'_{\mathrm c}(1)$ is logarithmically divergent:
\be
f'_{\mathrm c}(1)=- b \ln L_{\Gamma}\;
\label{f_c}
\ee
when $b=-C'_{\Gamma}(1)$. As a second point we study the universality of the prefactor, $b$.

\subsection{Entanglement entropy of the diluted quantum Ising model}
\label{sec:entropy}

The diluted quantum Ising model is defined by the Hamiltonian:
\be
{\cal H}=-\sum_{\left\langle ij\right\rangle } J_{ij} \sigma_i^x \sigma_j^x -\sum_i h \sigma_i^z\;,
\label{Ising}
\ee
in terms of the $\sigma_i^{x,z}$ Pauli matrices at site $i$. The first sum in Eq.(\ref{Ising})
runs over nearest neighbors and the $J_{ij}$ coupling equals $J>0$ with probability $p$ and
equals $J=0$ with probability $1-p$. At the percolation transition point, $p_c$, for small
transverse field, $h$, there is a line of phase transition the critical properties of
which are controlled by the percolation fixed point\cite{senthil_sachdev}. The ground state of ${\cal H}$
is given by a set of ordered clusters, which are in the same form as for percolation:
in each ordered cluster the spins are in a maximally entangled GHZ
state. Now let us consider the entanglement entropy, ${\cal S}_{\Gamma}$ between a subsystem and the environment, which are
separated by $\Gamma$. 
For a given realization of disorder in the small $h$ limit ${\cal S}_{\Gamma}$ is given
by the number of so called \textit{crossing clusters}, which intersect $\Gamma$ and contain also at least one point of the
environment\cite{lin07,yu07}. Evidently there are not more crossing clusters, than touching clusters,
thus ${\cal S}_{\Gamma} \le N_{\Gamma}$, and
$\left\langle {\cal S}_{\Gamma}\right\rangle \le \left\langle N_{\Gamma} \right\rangle$, however the corner contributions
for both quantities are expected to be asymptotically identical.
In the numerical calculations we shall consider both type of clusters and for simplicity instead of ${\cal S}_{\Gamma}$ we
use the wording $N_{\Gamma}$ of crossing clusters as well.

\section{Numerical results}
\label{sec:numerical}

We have performed large scale numerical calculations\cite{sedgewick} for site and bond percolations at the critical
point on the simple cubic lattice, with $p_c=0.2488126$ and $p_c=0.311608$, for bond and
site percolation, respectively\cite{perc_cr,note1}. The finite systems we have used have $L \times L \times L$ sites with $L$ up to
512 and with periodic boundary conditions. In the numerical calculations we have used different shapes
of the subsystems, which have either a $3d$ form: cube and sheared cube, or a lower-dimensional form: contour of a square or
sheared square and line segment. In all cases the linear size of $\Gamma$ is $L_{\Gamma} \sim L$ and
in Eq.(\ref{f_c}) we shall use $L$ instead of $L_{\Gamma}$.
For a given $\Gamma$ we have calculated the number of crossing and touching clusters, as defined in Sec.\ref{sec:entropy}, which are
then averaged:
i) for a given percolation sample over the positions of $\Gamma$ (typically $10^3$ positions), and
ii) over different samples. Typically we have used $10^4$ samples for each size, $L$, except the
largest ones, where we had at least $10^3$ samples. The corner contribution is calculated using the geometric approach
introduced earlier: for each sample $\left\langle N_{\Gamma} \right\rangle$ is calculated in three geometries (for $3d$
subsystems and contours) and in two geometries (for lower-dimensional contours) and by combining these data the corner contribution is
directly obtained for each sample.

\subsection{Cube subsystem}
\label{sec:cube}
\begin{figure}[!ht]
\begin{center}
\includegraphics[width=2.6in,angle=0]{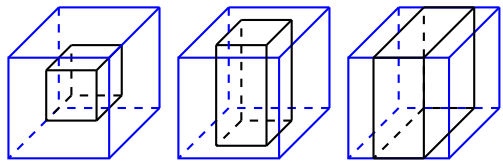}
\end{center}
\caption{
\label{fig_1} (Color online) Cube, column and slab geometries used in the calculation of the corner contribution of a cube
subsystem.}
\end{figure}

\begin{figure}[!ht]
\begin{center}
\includegraphics[width=3.2in,angle=0]{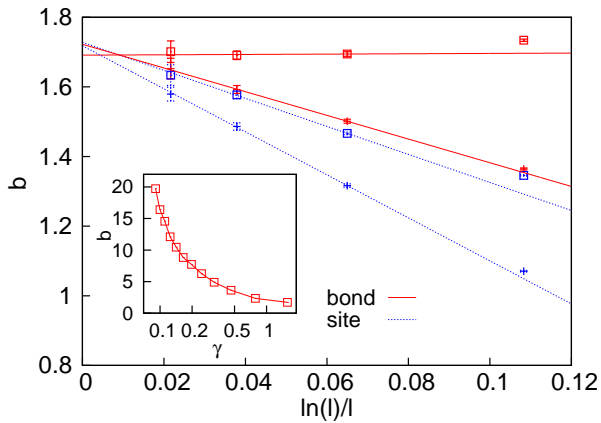}
\end{center}
\caption{
\label{fig_2} (Color online) Effective, size-dependent prefactors calculated for a cube subsystem for bond and site
percolation, as well as for touching ($\boxdot$) and crossing ($+$) clusters at $l=L/2$. The extrapolated value, $b=1.72(3)$ is universal. Inset:
Angle dependence of the extrapolated value of the prefactor for sheared cubes using touching clusters for bond
percolation (the error is smaller, than the symbol).
The lines across the symbols are guide to the eye.}
\end{figure}

The first geometry we consider for $\Gamma$ is a cube of linear size $L/2$. In the geometric
approach we used three different geometries, which are illustrated in Fig.\ref{fig_1}. The values of $N_{\Gamma}$
calculated in the different geometries are combined in a way, that the corner (as well as the surface and the edge)
contribution is obtained for each sample. This method is explained
in the Appendix of Ref.\onlinecite{kovacs_igloi12}. The numerical results show, that the corner contribution
of $\left\langle N_{\Gamma} \right\rangle$ has indeed a logarithmic size-dependence, as predicted in Eq.(\ref{f_c}),
but no (logarithmic) singular contributions are detected in the other terms in Eq.(\ref{N_Gamma}), such as in $f'_s(1)$
and in $f'_e(1)$.
Comparing the average corner contributions in finite systems of size $L$ and $2L$ we have
obtained effective, size-dependent estimates for the prefactor by two-point fits. The
results are presented in Fig.\ref{fig_2} both for site and bond percolations and for touching and crossing clusters.

\begin{figure}[!ht]
\begin{center}
\includegraphics[width=3.2in,angle=0]{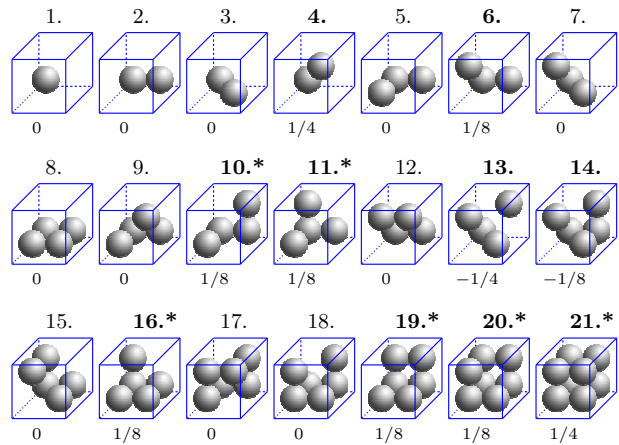}
\end{center}
\caption{
\label{fig_3} (Color online) Large scale topology of the clusters:
the cube is divided into eight identical parts and the possible
topologies of clusters in these parts are illustrated: empty box $\to$ no site; filled box $\to$ at least one occupied
site. Among the 21 distinct cluster geometries there are 10 with non-zero corner contribution, the value of which is written
below the configuration. For percolation only the connected clusters count indicated by an asterisk.}
\end{figure}

For large $L$ all these estimates approach the same universal value, $b=1.72(3)$.
As in 2d the corner contribution is related to the large scale topology of the clusters, which is illustrated
in Fig.\ref{fig_3}. As explained in the caption of this figure the average corner contribution, $N^{\mathrm{cr}}_{\mathrm{cube}}$, is
expressed as the sum of the relevant corner probabilities:
\beqn
N^{\mathrm{cr}}_{\mathrm{cube}}(L)&=&\frac{1}{8}\left[P_{10}(L)+P_{11}(L)+P_{16}(L)\right.\\ \nonumber \cr
&+&\left.P_{19}(L)+P_{20}(L)+2P_{21}(L)\right]\;,
\eeqn
where $P_i(L)$ refers to the $i$-th configuration in Fig.\ref{fig_3}. Note, that each of the non-vanishing
corner probabilities have a logarithmic size-dependence: $P_i(L) \simeq u_i \ln L$ with universal prefactors, $u_i$.

\begin{figure}[!ht]
\begin{center}
\includegraphics[width=3.2in,angle=0]{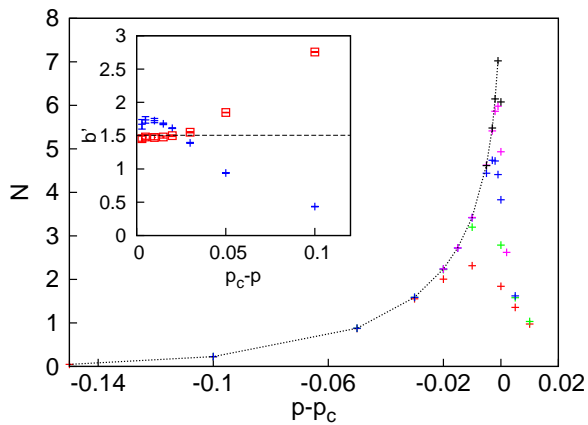}
\end{center}
\caption{
\label{fig_4} (Color online) $\Delta p=p-p_c$ dependence of the corner contribution to $N_{\mathrm{cube}}$ for bond percolation and
for crossing clusters for different sizes: $L=32,64,\dots,512$ from below. Close to $p_c$ the extrapolated curve has a logarithmic
singularity. In the inset the effective prefactors, $b'$ are shown as a function of $\Delta p$ for both types of clusters.
The straight dashed line represents the extrapolated value $b'=1.505(28)$, see in Eq.(\ref{S_corner_p}).}
\end{figure}

We have also studied the $p$-dependence of $N^{\mathrm{cr}}_{\mathrm{cube}}$ outside the critical point. As can be seen in Fig.\ref{fig_4}
$N^{\mathrm{cr}}_{\mathrm{cube}}$
has a peak around $p=p_c$ and close to $p_c$ the extrapolated curve can be well described by the scaling result:
\be
N^{\mathrm{cr}}_{\mathrm{cube}}(p) \simeq - b' \ln(p_c-p)+\mathrm{const.}\;,
\label{S_corner_p}
\ee
where $b'=b \nu$, with $\nu=0.875(1)$ being the correlation length critical exponent for $3d$ percolation\cite{perc_cr}. Indeed, assuming
the form in Eq.(\ref{S_corner_p}) we have calculated effective, $p$-dependent prefactors by two-point fits,
which are shown in the inset of Fig.\ref{fig_4}. The extrapolated value for $p \to p_c$
is consistent with scaling prediction.

We have also sheared the cube subsystem in a way, that one angle at the base is $\gamma < \pi/2$, while the other remains $\pi/2$ and the
volume of the system stays constant: $L^3/8$. For this system the geometric approach can be used for the specific values satisfying the
condition: $\tan(\gamma)=1/n$, with $n=0,1,\dots$ being an integer. The extrapolated numerical results about the prefactor of
the corner contributions are presented in the inset of Fig.\ref{fig_2}. $b(\gamma)$ is minimal at $\gamma=\pi/2$ and
for small $\gamma$ it is found to be divergent as $\sim 1/\gamma$.

\subsection{Square contour}
\label{sec:square}
\begin{figure}[!ht]
\begin{center}
\includegraphics[width=3.2in,angle=0]{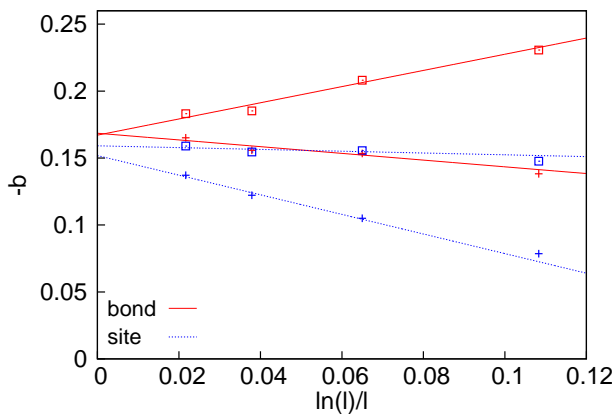}
\end{center}
\caption{
\label{fig_5} (Color online) Effective, size-dependent prefactors calculated for square contour for bond and site
percolation, as well as for touching ($\boxdot$) and crossing ($+$) clusters at $l=L/2$. The extrapolated value, $b=-0.158(10)$ is universal.}
\end{figure}

In the next example $\Gamma$ is the contour of a square of edge length $L/2$. The corner contribution in this case
is also logarithmically divergent at the critical point, but the prefactor is negative, in agreement with the considerations in
Ref.\cite{kovacs_igloi12}. Finite-size estimates for $b$ are presented in Fig.\ref{fig_5}. The extrapolated prefactor
is universal and found to be $b=-0.158(10)$, which is to be compared with the $2d$ result of $b=-5\sqrt{3}/(36 \pi)=-0.0766$.

\begin{figure}[!ht]
\begin{center}
\includegraphics[width=3.2in,angle=0]{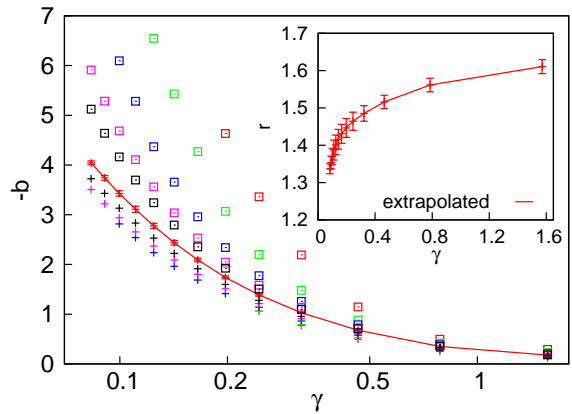}
\end{center}
\caption{
\label{fig_6} (Color online) Finite-size estimates of the prefactor for $L=32,64,\dots,512$ for touching clusters
($\boxdot$, from above) and for crossing clusters ($+$, from below) for sheared square contour as a function
of $\gamma$ (bond percolation). Inset: ratio of the extrapolated value of $b(\gamma)$ and the conformal result
for $2d$ percolation in Eq.(\ref{2d_sheared}): $r=b(\gamma)/(b_{2d}(\gamma)/\beta_{2d})$. The lines connecting the extrapolated
points are guide to the eye.}
\end{figure}

The square contour can be sheared as well, in a way that the smaller angle is $\gamma<\pi/2$, while the surface of the system
remains the same: $L^2/4$. The geometrical approach is used for the discrete values of $\gamma$-s, which are
listed in the previous section for the sheared cubes. The effective size-dependent prefactors of the corner contribution
are listed in Fig.\ref{fig_6} for bond percolation and for both types of clusters. The extrapolated value of
$b(\gamma)$, which is also indicated in Fig.\ref{fig_6} is found to be the same for site percolation within the
accuracy of the calculation; it is increasing with decreasing $\gamma$, being divergent as $\sim 1/\gamma$
for small $\gamma$. This behavior is similar to that in $2d$, therefore we have compared the extrapolated $b(\gamma)$ with the conformal
result in $2d$: 
\be
b_{2d}(\gamma)=-\frac{\beta_{2d}}{12}\left[4-\pi\left(\frac{1}{\gamma}+\frac{1}{\pi-\gamma}+\frac{1}{\pi+\gamma}+\frac{1}{2\pi-\gamma}\right)\right]\;.
\label{2d_sheared}
\ee
with $\beta_{2d}=5 \sqrt{3}/(4 \pi)$. In the inset of Fig.\ref{fig_6} we present the ratio $r=b(\gamma)/(b_{2d}(\gamma)/\beta_{2d})$,
which has a considerable $\gamma$-dependence.

\subsection{Line segment and crosses}
\label{sec:line}
One of the simplest contours are the straight line segment of length $L/2$ and crosses of straight line segments.
For the latter we consider $n=1,2,3$ or $4$ crosses, which lay in the same $2d$ plane of the simple cubic lattice.

\begin{figure}[!ht]
\begin{center}
\includegraphics[width=3.2in,angle=0]{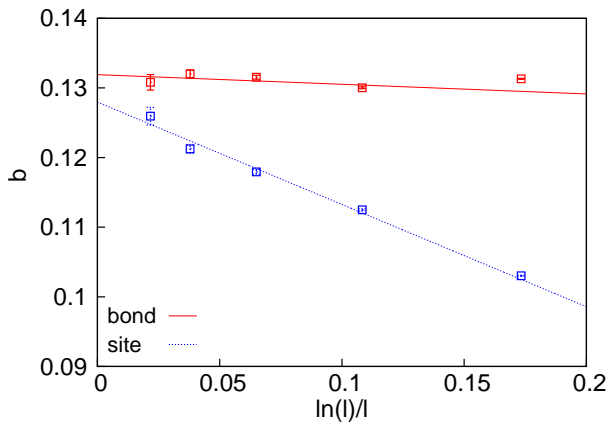}
\end{center}
\caption{
\label{fig_7} (Color online) Effective, size-dependent prefactors calculated for the line segment for bond and site
percolation at $l=L/2$. The extrapolated value, $b=0.130(3)$ is universal.}
\end{figure}

For the line segment the geometric approach is described in Ref.\cite{kovacs}. The calculated finite-size estimates for the prefactor of
the corner contribution are presented in Fig.\ref{fig_7}. The extrapolated value, $b=0.130(3)$ is found to be universal, which is
to be compared
with the result in $2d$: $b_{2d}=5\sqrt{3}/(32 \pi)=0.08615$.

For crosses the calculation of $N_{\Gamma}$ has been introduced in Ref.\cite{kovacs14}, and we refer to Fig.2 in this paper for the
representation of $n=1,2,3$ and $4$ crosses. In these cases the total corner contribution is given
as the sum of two type of individual contributions: i) being at the end of a line (with a contribution ${\cal B}_{end}$) and
ii) being at the crossing point of two lines (with a contribution ${\cal B}_{cross}$).
If we normalize the total corner contribution to one cross, than it can be expressed as: $4{\cal B}_{end}+{\cal B}_{cross}$,
$2{\cal B}_{end}+{\cal B}_{cross}$,$4/3{\cal B}_{end}+{\cal B}_{cross}$ and ${\cal B}_{cross}$,
for $n=1,2,3$ and $4$ crosses, respectively. Note, that the line segment has two end corners, thus the prefactor
equals to $2{\cal B}_{end}$. All the results
are found to be consistent with the estimates: ${\cal B}_{end}=0.0650(15)$ and ${\cal B}_{cross}=-0.4044(64)$.

\section{Discussion}
\label{sec:disc}
In this paper we considered the mean number $N_\Gamma$ of clusters in $3d$ critical percolation which intersect
a contour $\Gamma$ in the cases when it has sharp corners or end points. When $\Gamma$ represents
the interface between a subsystem and the environment, then $N_\Gamma$ is related to the entanglement entropy of the diluted
quantum Ising model. We have shown by accurate numerical calculations, that $N_\Gamma$ has universal logarithmic terms, the
prefactors of which are the sum of the individual corner contributions. This result is analogous to the $2d$
behavior, in which case the logarithmic corner contributions are related to the free-energy singularity
of critical $2d$ systems, which follows from the Cardy-Peschel formula\cite{cardypeschel}. For recent numerical studies
in $2d$ systems see Refs.[\onlinecite{2dnum}]. In $3d$ there are no field-theoretical
conjectures about logarithmic free-energy singularities, but in the frame of phenomenological scaling theory\cite{privman}
similar singular behavior is expected in $2d$ and in $3d$. Using this argument in our case, the average cluster number
in Eq.(\ref{N_Gamma}) is written as the sum of a singular part:
\be
\left\langle N_{\Gamma} \right\rangle_s(\Delta p,L_{\Gamma})=Y(\Delta pL_{\Gamma}^{1/\nu})=-\frac{b}{\epsilon}+y(\Delta pL_{\Gamma}^{1/\nu})
+\dots\;,
\label{sing1}
\ee
and a nonsingular part: $\left\langle N_{\Gamma} \right\rangle_{\mathrm{ns}}(\Delta p,L_{\Gamma})$. The latter has a nonsingular corner
term in dimension $d+\epsilon$:
\be
\left\langle N^{\mathrm{cr}}_{\Gamma} \right\rangle_{\mathrm{ns}}(\Delta p,L_{\Gamma})=L_{\Gamma}^{\epsilon}\psi_{\mathrm{cr}}(\Delta p)=
L_{\Gamma}^{\epsilon}\left[ \frac{b}{\epsilon}+\overline{\psi}_{\mathrm{cr}}(\Delta p)+\dots \right]\;.
\label{sing2}
\ee
In Eqs.(\ref{sing1}) and (\ref{sing2}) the leading terms are expected to have divergent amplitudes, and the sum of those in
the $\epsilon \to 0$ limit leads to the logarithmic term in Eq.(\ref{f_c}). Also this interpretation is in agreement with the observed
universality of the prefactor $b$.
Our numerical studies reveal the existence of these logarithmic terms, and
to our knowledge our result is the first demonstration of the existence of such singular terms in a $3d$ system. Although our
investigations are limited to percolation, which is the $Q \to 1$ limit of the $Q$-state
Potts model, similar logarithmic terms are expected to exist for another (not necessary integer) values of $Q$, such as for the Ising
model with $Q=2$, as far as the transition is of second order. For general values of $Q$ the mean cluster numbers of Fortuin-Kasteleyn
clusters should be considered.

Our study is related to recent investigations of the entanglement entropy of the critical random ferromagnetic quantum Ising
models\cite{kovacs_igloi12}. According to the so called strong disorder renormalization group\cite{im} study the ground state of
this system (in any finite
dimension) is represented by a set of clusters, which are however, generally not connected, and this makes an important
difference with that of the ground state of the diluted quantum Ising model. The entanglement entropy of the random model is also
related to the number of crossing clusters at the interface, and it has also a logarithmic corner term, although its prefactor
is different from that in the diluted model. This behavior has been established in the framework of the strong disorder
renormalization group method. Interestingly, for end points, as for a line segment, the random model has no logarithmic
corrections, while the dilute model do has. Since the line segment is perhaps the simplest geometry, we expect that the observed
logarithmic corner contribution to $N_\Gamma$ for critical $3d$ percolation could be proven by some analytical method, such as
by field theory.

\begin{acknowledgments}
This work has been supported by the Hungarian National Research Fund under grant
Nos. OTKA K75324, K77629 and K109577. The research of IAK was supported by
the European Union and the State of Hungary, co-financed by the European Social Fund in the framework
of T\'AMOP 4.2.4. A/2-11-1-2012-0001 'National Excellence Program'. The authors thank to John Cardy, Eren Metin El\c{c}i and
Martin Weigel for previous cooperation in the project.
\end{acknowledgments}


\begin{thebibliography}{99}
\vskip -.5cm

\bibitem{stauffer} D. Stauffer and A. Aharony, \textit{Introduction to percolation theory} (2nd ed.), Taylor \& Francis, London, 1992.

\bibitem{smirnov} S. Smirnov, C. R. Acad. Sci. Paris S\'er. I Math., 333 , no. 3, 239-244 (2001);
J. Tsai, S.C.P. Yam, W. Zhou, arXiv:1112.2017.

\bibitem{dotsenko_fateev} Vl.S. Dotsenko and V.A. Fateev, Nucl. Phys. B \textbf{240}, 312 (1984).

\bibitem{crossing} J.L. Cardy, J. Phys. A\textbf{25}, L201 (1992).

\bibitem{conf_inv} J.L. Cardy, in \emph{Phase Transitions and Critical Phenomena}, edited by C. Domb
and J.L. Lebowitz (Academic Press, London, 1987), Vol. 11. p. 55.

\bibitem{sle} O. Schramm, Israel J. Math. 118, 221 (2000);
S. Smirnov and W. Werner, Math. Research Letters 8, no. 5-6, 729-744 (2001).

\bibitem{kovacs} I. A. Kov\'acs, F. Igl\'oi and J. Cardy, Phys. Rev. B 86, 214203 (2012).

\bibitem{cardypeschel} J. Cardy and I. Peschel, Nucl. Phys. B, 300 [FS22],
377 (1988).

\bibitem{kovacs14}I. A. Kov\'acs, E. M. El\c{c}i, M. Weigel and F. Igl\'oi,
Phys. Rev. B\textbf{89}, 064421 (2014).

\bibitem{privman_fisher} V. Privman and M.E. Fisher, Phys. Rev. B\textbf{30}, 322 (1984).

\bibitem{privman} V. Privman, Phys. Rev. B\textbf{38}, 9261 (1988).

\bibitem{lin07} Y-C.Lin, F. Igl\'oi and H. Rieger, \prl \textbf{99}, 147202 (2007).

\bibitem{yu07} R. Yu, H. Saleur and S. Haas, \prb \textbf{77}, 140402 (2008).

\bibitem{wu} F.Y. Wu, Rev. Mod. Phys. 54, 235 (1982).

\bibitem{Fortuin-Kasteleyn} P. W. Kasteleyn and C. M. Fortuin, J. Phys. Soc. Japan \textbf{26}, 11 (1969).

\bibitem{touching} These clusters are called as \textit{touching clusters}, while the \textit{crossing
clusters} are defined in Sec.\ref{sec:entropy}.

\bibitem{senthil_sachdev} T. Senthil and S. Sachdev, Phys. Rev. Lett. \textbf{77}, 5292 (1996).

\bibitem{sedgewick} We applied the weighted union-find algorithm with path compression, see for example R. Sedgewick, {\sl Algorithms}, 2nd edition, Addison-Wesley, Reading, Mass. (1988).

\bibitem{perc_cr} C. D. Lorenz, and R. M. Ziff, Phys. Rev. E\textbf{57}, 230 (1998); J. Phys. A \textbf{31} 8147 (1998).

\bibitem{note1} After performing most of our numerical calculations we have noticed a more recent estimate for the critical point
of $3d$ percolation: $p_c=0.24881182(10)$ (bond) and $p_c=0.3116077(2)$ (site) by J. Wang \textit{et al}, Phys. Rev. E \textbf{87},
052107 (2013). We have checked that using this values for $p_c$ would modify our estimates in a negligable extent: the difference
being much smaller than the statistical error.

\bibitem{kovacs_igloi12} I. A. Kov\'acs and F. Igl\'oi, EPL \textbf{97}, 67009 (2012).

\bibitem{2dnum} X. Wu, N. Izmailian, W. Guo, Phys. Rev. E \textbf{86}, 041149 (2012); \textit{ibid} \textbf{87}, 022124 (2013);
N. Izmailian, R. Kenna, W. Guo, X. Wu, arXiv:1402.5856.


\bibitem{im} For a review, see: F. Igl\'oi and C. Monthus, Physics Reports {\bf 412}, 277, (2005).
\end{thebibliography}
\end{document}